# Personalised network modelling in epilepsy

Yujiang Wang [a,b,c], Gabrielle Marie Schroeder[ a], Nishant Sinha [a,b],
Peter Neal Taylor[a,b,c*]

Epilepsy is a disorder characterised by spontaneous, recurrent seizures. Both local and network abnormalities have been associated with epilepsy, and the exact processes generating seizures are thought to be heterogeneous and patient-specific. Due to the heterogeneity, treatments such as surgery and medication are not always effective in achieving full seizure control and choosing the best treatment for the *individual* patient can be challenging. Predictive models constrained by the patient's own data therefore offer the potential to assist in clinical decision making.

In this chapter, we describe how personalised patient-derived networks from structural or functional connectivity can be incorporated into predictive models. We focus specifically on dynamical systems models which are composed of differential equations capable of simulating brain activity over time. Here we review recent studies which have used these models, constrained by patient data, to make personalised patient-specific predictions about seizure features (such as propagation patterns) or treatment outcomes (such as the success of surgical resection).

Finally, we suggest future research directions for patient-specific network models in epilepsy, including their application to integrate information from multiple modalities, to predict long-term disease evolution, and to account for within-subject variability for treatment.

**Affiliations**
a Interdisciplinary Computing and Complex BioSystems Group, School of Computing, Newcastle University, UK
b Institute of Neuroscience, Faculty of Medical Science, Newcastle University, UK
c Institute of Neurology, University College London, UK
* peter.taylor@newcastle.ac.uk

# 1. Introduction

Epilepsy is a family of neurological disorders in which patients experience unprovoked spontaneous seizures. Unfortunately, there is currently no cure for epilepsy, and seizure management is the target of most therapies. The first-line treatment of epilepsy is usually anti-epileptic drugs. However, depending on the subtype of epilepsy and the individual, drug treatments fail to control the seizures in around one-third of patients. One challenge in the treatment of epilepsy is its heterogeneity. In each patient, seizures are thought to be generated by different mechanisms/processes/parameters, and treatment outcomes will also depend on these.

Fortunately, epilepsy is one of the disorders that can be captured and measured on many spatial and temporal scales, even in human patients. For example, at the micro-scale, it is possible to infer single neuron firing patterns in patients using microelectrodes (Truccolo *et al.* 2011, Merricks *et al.* 2015). At a coarser spatial scale, scalp EEG and intracranial EEG allow long-term recording of spatio-temporal electrical signals (Cook *et al.* 2013). Concurrent measurement of electrical signals with blood oxygenation at the whole brain scale using functional MRI are also possible (Coan *et al.* 2016), even during seizures (Möller *et al.* 2008); and structural measurements of white-matter integrity and connectivity can also be made using diffusion MRI (Duncan *et al.* 2016). High prevalence rates additionally enable recruitment of large cohorts for studies, and the lifetime impact of the disease facilitates longitudinal analyses (Lossius *et al.* 2008). This wealth of data enables the development of data-driven approaches to understanding patient-specific disease mechanisms, and designing patient-specific treatment strategies.

A key component in achieving data-driven and patient-specific treatments are personalised models of the epileptic dynamics in each patient. Models are generally useful in situations where we wish to predict something that is unknown, such as a treatment outcome, an impending seizure, or a parameter that cannot be directly measured. Personalised models usually make use of a general model structure (e.g. reflecting some basic assumptions on the structure and function of the brain), and additionally use patient-specific data to parameterise and validate it. Such personalised models can then be used to make predictions about disease mechanisms or treatment outcomes in individual patients.

1.1. Dynamical systems models and incorporation of personalised brain connectivity

Many types of models exist, including conceptual models based on a pre-conceived hypothesis, machine learning models built on data-driven features, and dynamical systems models consisting of parameters and variables. Dynamical systems models are particularly useful for simulating time-varying systems, especially when some prior knowledge of the mechanisms of the system are known. They have recently shown promise for making personalised predictions in epilepsy, and are the focus of this chapter.

A dynamical systems model is composed of a set of equations which usually describe how a property of interest changes with respect to time. A classic example in the neurosciences is the neural population model of Wilson and Cowan (1972), where the equations describe how neural population firing rates change over time. In that model two equations (variables) are used to represent excitatory and inhibitory neural populations. The assumption of the excitatory and inhibitory populations represents a prior knowledge of the system. A set of parameters and nonlinear functions govern all of the ways in which the two populations can possibly interact. More complex models can also be designed, incorporating additional, more specific populations of neurons, such as the Jansen-Rit (1995) model. The additional complexity enables the model to have different types dynamics, but at the cost of requiring more assumptions about the populations' behaviours and interactions.

Dynamical systems models of seizures usually use such neural population models to simulate the brain dynamics seen on EEG, as the EEG is thought to capture brain dynamics arising from populations of neurons. In the model, high amplitude oscillatory activity is commonly used to represent seizure (ictal) dynamics, while low amplitude irregular dynamics are understood as the non-seizure (interictal) state. Most earlier work investigated mechanisms of seizure onset in the model (da Silva *et al.* 2003). Seizure onset mechanisms have typically been modelled either as a parameter change through a bifurcation or as a noisy process in a bistable system, although other mechanisms are possible (Baier *et al.* 2012). Parameter-induced seizure transitions work by simulating a slow change in a parameter such that the non-seizure state ceases to exist and the only stable attractor is a seizure state. The parameter value for which the state change occurs is known as a bifurcation point. The alternative approach uses bistability, where the seizure and non-seizure states both coexist, and external input drives seizure transitions. Excitability parameters can then be defined as the proximity to the bifurcation (Wang *et al.* 2012), or likelihood of transition in a bistable system (Hutchings *et al.* 2015).

Traditionally, most previous epilepsy modelling work at the macroscopic spatial scale has not incorporated complex spatial interactions or personalised patient data into the models (Taylor *et al.* 2014). This approach has only happened relatively recently, and has been partly been driven by the increased understanding of how network interactions contribute to epileptic processes, even in focal epilepsies (Richardson 2012). To incorporate patient networks into the model parameters, a set of equations which captures one of the above seizure-transition mechanisms is then coupled to another set of equations. The coupling typically takes the form of a connectivity matrix inferred from patient data. Simulated seizures can then occur in the model, and their spatio-temporal appearance is dependent on the model and its (patient-specific) connectivity parameters. Figure 1 shows a schematic of the approach.

In this chapter, we will discuss some recent examples which incorporate patient-specific connectivity into computational models for predictive value. This chapter is structured as follows: In part 2 we discuss how subject-specific structural connectivity has been used to constrain model parameters. In part 3 we discuss examples where patient-specific EEG functional connectivity has been used to constrain model parameters. Finally, we suggest some possible future directions in part 4.

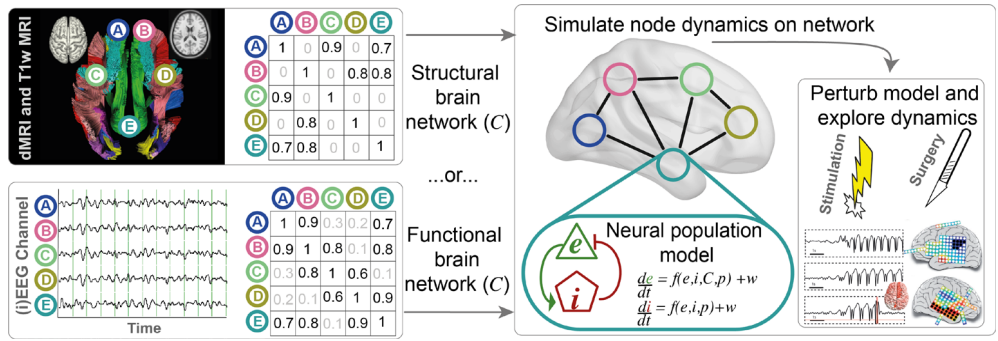

**Figure 1: Schematic of processing and analysis pipeline for personalised brain network simulations.** First, a personalised brain network is generated which in this example contains nodes (A-E) and connections between them (entries in the matrix). Connections can be inferred from diffusion weighted MRI tractography (upper left panel), or from associations between time series recordings (lower left panel). This personalised brain network is then used for simulations, by placing a neural population unit in each node. The coupling between the nodes is dictated by the personalised brain network. Simulations result in time series for each variable (typically at least two per node – e.g. one excitatory and one inhibitory). The simulated time series can be analysed to make inferences about the influence of the personalised connectivity on the dynamics. Perturbations or experiments can also be performed, e.g. by stimulating the model (through a simulated injection of current to one or more variables and nodes), or simulating surgery (by disconnecting nodes and re-simulating the dynamics). Insets show examples of simulated personalised stimulation and surgery (Taylor *et al.* 2015, Sinha *et al.* 2017).

## 2. Structural connectivity-based modelling

Structural whole-brain networks (connectomes) represent different brain regions that are interconnected by white matter fibre tracts. These networks are typically reconstructed from diffusion weighted magnetic resonance imaging (dMRI). From these images, white matter tracts reflecting the connectivity between different brain regions can be inferred using a fibre tracking algorithm. A network formulation is then possible by parcellating grey matter into predefined regions of interest, also referred to as nodes, with the fibre tracts connecting them becoming the edges (see Hagmann *et al.* 2008 for an early example). Though dMRI tractography is not usually routinely analysed clinically in epilepsy, recent studies have illustrated that it allows clinically useful predictions (Winston *et al.* 2011, 2012, Taylor *et al.* 2018) and may improve understanding of the pathophysiology of epilepsy. For example, diffusion MRI has revealed reduced fractional anisotropy in temporal lobe epilepsy (Ahmadi *et al.* 2009; Concha *et al.* 2005) and other structural network alterations in epilepsy (Bonilha *et al.* 2012; Besson *et al.* 2014; *DeSalvo et al.* 2014). However, static brain networks have limited ability to explain emergent functional and pathological spatiotemporal dynamics, such as seizures.

Dynamical systems models, when combined with patient-specific structural connectomes, enable the construction of personalised models that can simulate possible spatiotemporal dynamics arising from each network structure. Importantly, they can also simulate spatiotemporal dynamics arising from *changes* to patient-specific connectomes (e.g. due to surgery or disease progression). When validated, these models can then be applied to predict treatment outcome or design treatment strategies.

Some of the early work simulating brain dynamics based on changes to structural connectomes was not in the context of epilepsy. For example, Honey *et al.* (2008) illustrated the effect of cortical lesion induced changes in simulated brain dynamics using the macaque structural brain connectivity network as the coupling parameter in two different models. Lesioning was performed by removing connections to/from specific nodes, and model outputs compared pre- and post-lesioning. The authors showed that the effect on cortico-cortical interactions due to lesioning the high degree nodes was most widespread. Community architecture was highlighted as a significant predictor in determining whether the dynamical consequences of lesions should remain confined to a cluster. This work illustrated a framework for how model simulations, combined with structural connectomes, could be used to study the behaviour arising from a static structural network.

In the context of epilepsy, some publications used the framework above to study the effect of epilepsy surgery. For example, in an exploratory study, Hutchings *et al.* (2015) used a bistable phenomenological model for each node (brain region) that could transition between non-seizure and seizure-like states. The coupling parameters were derived from the dMRI from patients with temporal lobe epilepsy. Node excitability parameters were set higher in more atrophied areas in a data-driven manner using healthy control volumetrics as a reference – i.e. the excitability was proportional to the abnormality of the brain region. Upon simulation, the model exhibited transitions to seizure-like states more often using the patient parameters than those of the healthy controls. Notably, the amygdala, hippocampus, and parahippocamphal gyrus - which are routinely removed clinically - were also predicted amongst the most seizure prone by the model. Surgical outcomes were also predicted by simulating the changes in transition time due to given (planned) clinical resections compared with the resection performed in random areas. A limitation of that study was that the simulated surgeries did not account for individual variations in surgery (Taylor *et al.* 2018), nor was the model validated with personalised patient outcomes. However, it did demonstrate the group level finding of wide variations in post-operative surgical changes in seizure likelihood between patients, similar to those observed clinically (De Tisi *et al.* 2011).

A more recent study by Jirsa *et al.* 2017 used the 'Epileptor', a phenomenological model comprising of two oscillators coupled by a slow variable. In the model, transitions were simulated though saddle node and homoclinic bifurcations at seizure onset and offset, respectively. These two types of bifurcations had previously been shown to replicate features of seizure onset and offset well (Jirsa *et al.* 2014). The brain network model was designed by placing the Epileptor at each network node and coupled to each other using patient's diffusion MRI-inferred connectome. Excitability parameters of local nodes were set according to if the node was deemed epileptogenic zone or propagation zone based on clinical assessment. The parameter space was then explored for simulating network seizure dynamics. Extending the modelling framework in Jirsa *et al.* 2017, seizure propagation was modelled by Proix *et al.* 2017. Specifically, the hypothesised epileptogenic zone was quantified from neuroimaging data (MRI and intracranial EEG) collected during the presurgical evaluation of intractable epilepsy patients. Setting the parameter of nodes in the hypothesised epileptogenic zone to be more excitable, linear stability analysis was applied to identify nodes in the propagation zone. The model was validated by computing the accuracy of the predicted propagation zone with its clinical estimates. The authors further showed a correspondence between the surgical outcomes and the proportion of propagation zone outside the coverage of invasively placed intracranial EEG electrodes - i.e., the number of unexplored regions in the propagation zone was significantly higher in poor surgical outcome group. This suggests that in those patients who were not-seizure-free, the epileptogenic and/or propagation zone was not fully captured by iEEG. Obvious clinical applications would be to use this approach to guide electrode placement, as well as to predict post-operative seizure freedom.

Proix *et al.* 2018 then used the Epileptor in a multi-scale approach to model spatiotemporal patterns of seizure spread and termination. The Epileptor model was extended to a neural field model in which the short-range connectivity was incorporated by homogeneous coupling and long-range connectivity with heterogeneous coupling. The interplay between spatial and temporal scales in the model explained two observations of spike-wave discharge (SWD) type focal seizures: (i) propagation of slow ictal wavefront and fast SWDs (ii) asynchronous termination of seizures in clusters. This study suggested that these dynamics, together with variations in short- and long-range connectivity strength, play a central role in seizure spread, maintenance and termination.

In contrast to the studies above, which simulate focal onset seizures, other studies have performed personalised brain network modelling of patients with generalised seizures. Yan & Li (2013) investigated the role of structural connectivity in inducing abnormal hypersynchrony during seizures. There, the anatomical connectivity and conduction delay between different cortical areas were inferred from healthy subjects' dMRI. Modelling different cortical areas using a coupled system of four Kuramuto oscillators at each area, the authors investigated the role of connectivity and delays on large-scale synchronisation. Importantly, the model quantified the degree of synchronisation of each cortical area locally as well as the whole cortex globally. It was apparent that some cortical areas ("hot spots") have a higher propensity of inducing global synchronisation than others; this is not otherwise observable on the recordings of SWD electroencephalogram. To our knowledge this was the first study to simulate dynamics on a human dMRI structural network in the context of epilepsy. Even though the networks used were derived from healthy subjects, the same approach could be applied using patient data.

In another early study of generalised epilepsy modelling, Taylor *et al.* (2013) also used a multi-scale approach to simulate between-subject variations in generalised spike-and-wave seizure dynamics. There, four neural population models were coupled locally to represent the dynamics within an individual node/brain region. Whole-brain scale long-range connections between regions were then inferred from healthy subjects' dMRI. In other words, different "patients" were simulated, although the connectome underlying each patient's simulation was derived from healthy subjects. Using this multi-scale approach, the authors demonstrated that the model can reproduce the prototypic waveform of spike-and-wave discharges (SWDs). Simulated electrographic data closely resembled the clinical recordings of absence seizures. The simulated SWD seizure dynamics showed higher within-"patient" similarity than between "patients", thus hypothesising that variations in SWD dynamics between patients may be due to their personal connectome. This modelling approach facilitated development of an optimal stimulation algorithm for abating seizures *in silico* where the stimulation was tailored to the individual (Taylor *et al.* 2015), suggesting a potential therapeutic method towards seizure control in generalised epilepsy. A limitation of both the early studies (Yan & Li 2013, Taylor *et al.* 2013) is that they used connectivity derived from healthy control imaging; however, the same approach could be used to simulate emergent dynamics on patient-specific structural networks. This short-coming was partly addressed in the study by Taylor *et al.* (2015), which used dMRI acquired from a patient with idiopathic generalised epilepsy.

Two other studies have applied dMRI-informed models to conceptually different questions in the context of epilepsy. In the first study, Lu *et al.* (2018) used connectivity from healthy subjects' dMRI, combined with a neural mass model to simulate EEG and functional MRI. The authors used the model to generate music from a simulated seizure. They suggest that the "level of arousal" while participants listen to the music could be used as a potential tool to discriminate patient populations. In a second study, Abdelnour *et al.* (2015) modelled the atrophy pattern in mesial temporal lobe epilepsy using two network-based models: (i) propagation of epileptogenic activity (i.e. extra-hippocampal spread of seizure activity), and (ii) progressive neurodegenerative process whereby loss in hippocampal neurons leads to loss in other connected regions. Both network models simulate a diffusion process on the whole-brain structural connectivity network. The later model of atrophy spread reproduced the empirically observed atrophy significantly better than the former. Although the structural connectivity was acquired from healthy controls, atrophy measurements were derived from patients. These models might be helpful in predicting future spread of atrophy which may pave way for a tailored surgical therapy to prevent progressive degeneration.

### 3. Functional connectivity-based modelling
While determining structural connectivity helps reveal alterations in brain structure in epilepsy, it is also important to understand the changes in neural dynamics that occur in this disorder. In particular, the high temporal resolution of electroencephalographic (EEG) recordings allows clinicians to identify abnormal patterns of activity during both interictal and ictal periods. The easiest recording to obtain is scalp EEG, where electrodes placed on the scalp record neural activity generated by the underlying cortex. However, in order to finely localize the origin of seizures in candidates for resective surgery, a more localized recording is required. Some patients therefore undergo intracranial EEG (iEEG) monitoring, in which contacts are surgically placed either directly on the surface of the brain (typically in grid or strip arrangements), or into deeper structures using depth electrodes (Javidan 2012, Cook *et al.* 2013, Taussig *et al.* 2015). Due to the invasive nature, iEEG provides a finer spatial resolution and less noisy recording than scalp EEG; and can also be used to observe the dynamics of deep brain regions such as the hippocampus. The overall spatial coverage provided by iEEG, though, is more limited; in each patient, only certain regions are recorded from based on the hypothesised epileptogenic zone.

Section 2 described computational models in which brain regions were coupled based on patient-specific structural connectivity. Likewise, the coupling parameters of such network models can be defined by functional connectivity, derived from scalp or intracranial EEG (Sinha *et al.* 2014, Goodfellow *et al.* 2016, Sinha *et al.* 2017, Lopes *et al.* 2018, Yang *et al.* 2018). In this case, each node represents the brain region recorded from by a single electrode, while edges correspond to the statistical relationship (e.g., correlation or coherence) between each pair of EEG signals. While structural connections provide opportunities for neural interactions, functional connectivity describes the actual changes in measured interactions during different brain states, such as seizures. Many studies have used functional connectivity to elucidate how spatiotemporal brain dynamics change before, during, and after seizures (Khambhati *et al.* 2016, Burns *et al.* 2014, Kramer *et al.* 2010), and the epileptogenic zone also appears to have unique functional network properties (Burns *et al.* 2014). As such, computational models incorporating patient-specific functional networks can explore how functional interactions may contribute to epileptic mechanisms in each patient.

Studies have used scalp EEG recordings to infer functional connectivity and applied them in the context of modelling generalised seizures. For example, the pioneering study by Benjamin *et al.* 2012, used a bistable model to investigate the impact of noise and network structure on seizure-like transitions in a model constrained by scalp EEG. There, the authors showed that the model produced significant differences between patients and controls in terms of the model output. Further work from the same group suggested the potential for EEG based modelling to discriminate between patients and controls (Schmidt *et al.* 2014). Finally, the study by Taylor *et al.* (2013b) used correlations between EEG channels to constrain a network model of the thalamocortical loop generating generalised spike-and-wave seizures. The authors further examined the effect of stimulation in this model, and found a spatial heterogeneity in terms of the stimulation response duration. This heterogeneity was suggested to account for spatial variability observed after stimulation in experimental models of spike-and-wave seizures (Zheng *et al.* 2012).

Meanwhile, models based on iEEG data from patients with focal epilepsy have focused on understanding and predicting the effects of surgical resection. Sinha *et al.* (2014) created patient-specific computational models based on interictal iEEG functional connectivity in a cohort of six patients with focal epilepsy, an approach that was later extended to a study of 16 patients (Sinha *et al.* 2017). In their study, each node of the network is a bistable model at the cusp of a subcritical Hopf bifurcation, and the nodes are coupled by the patient's interictal functional connectivity. Noise and external inputs have the potential to push each node into an oscillatory state, reminiscent of epileptic discharges. To quantify each model's seizure likelihood, the authors measure the "escape time" of each node – the amount of time it takes for the node to switch from a resting state to the oscillatory dynamics. In good outcome patients, removing the resected nodes from the model increases escape time (i.e., decreases seizure likelihood) more than a random resection. Meanwhile, alternative resection for bad-outcome patients are proposed based on which nodes have the highest seizure likelihood. The authors suggest the modelling approach for predicting the outcome of surgery based on interictal data and suggesting alternative resection strategies.

Using ictal iEEG data, Goodfellow *et al.* (2016) simulated dynamics on functional networks using a separate cohort of sixteen patients with focal epilepsy who underwent iEEG monitoring and surgical resection. They modelled the dynamics at each node using a neural mass model (Wendling *et al.* 2002) set close to a saddle-node on invariant circle bifurcation. Coupling between nodes was proportional to patient-specific functional networks, which were derived from the first half of seizure activity. Because the node dynamics are close to a bifurcation, noisy inputs produce simulated discharges. They define "Brain Network Ictogenicity" (BNI) as the proportion of time the model spends in discharges, thus allowing them to quantify the effects of perturbations on the model dynamics. First, they observed that if the resected nodes are removed from the network, BNI is reduced more for patients that had good surgical outcomes; in other words, successful interventions had a greater impact on model dynamics than unsuccessful surgeries. They then proposed alternative resections based on which nodes produced the greatest reduction in discharges upon removal, and found their proposed resections were consistent with the actual resections of good, but not bad, outcome patients. In a later study, Lopes *et al.* (2018) used the same concept of BNI to explore how patient-specific network dynamics change close to and during seizures. They demonstrated that surgical outcome predictions are more reliable for patients that consistently have an increase in BNI during seizures, compared to pre- and post-ictal periods. They propose that in these patients, the ictogenic network is well-represented by the patient-specific functional connectivity, which can therefore be used to devise an optimal resection strategy.

Despite some differences in their approaches, these studies all suggest that modelling dynamics on patient-specific functional networks can identify the drivers of abnormal neural activity, predict the outcome of a proposed resection, and suggest both more effective and less extensive surgical targets (Goodfellow *et al.* 2016, Sinha *et al.* 2017). While early work focused on static network representations of ictal (Goodfellow *et al.* 2016) or interictal (Sinha *et al.* 2017) activity, there has been recent interest in exploring how changing functional interactions influence model behaviour (Lopes *et al.* 2018, Yang *et al.* 2018). Interestingly, in all of these models, the intrinsic dynamics of each node are assumed to be the same – it is solely the model network structure that produces differences in the nodes' dynamics. This observation is consistent with theoretical studies demonstrating that network structure, rather than node properties, can be the primary source of abnormal dynamics (Hebbink *et al.* 2017). However, this is in contrast to some previously mentioned approaches using structural connectivity, where both intrinsic node dynamics and network structure give rise to the model behavior (Hutchings *et al.* 2015, Proix *et al.* 2017). A recent review further highlights this question of node vs. network cause of focal seizures (Smith & Schevon, 2016). Personalised network modelling approaches using structural and functional connectivity may prove crucial in future to answer this question.

## 4. Opportunities and future applications

The studies highlighted above suggest the potential benefits of a personalised modelling approach in the context of epilepsy; however, there are still some challenges to overcome in order for the approach to be useful for real-world clinical applications. Here, we highlight some areas which we consider to be unsolved challenges or underdeveloped research areas thus far. We suggest they may serve as fruitful avenues of future investigation.

4.1 Multimodal data integration
Currently, personalised models are usually created with one patient-specific component (e.g. the patient's structural brain network) and validated against a variable, such as surgical outcome. However, in the era of multimodal data, it is unclear how to best include all the available information into one personalised model. One excellent example of how this challenge may be tackled is given by Proix et al. (2017). On top of the patients' structural connectivity, Proix *et al.* (2017) additionally used EEG derived information to further parametrise the model (in terms of local excitability of each node), and predict the propagation of ictal activity beyond the recording sites. The validation of those predictions with another data modality - the actual patient outcomes - serves as an example of how the field can progress.

Another possibility is to consider multiple modalities for the validation of a model. For example, the study by Proix *et al.* predicted the existence and location of possible epileptogenic tissue in areas not recorded by iEEG. Their validation was essentially patient outcomes. However, additional validation from e.g. source localised MEG, high density scalp EEG, or fMRI could also be used to investigate the presence of abnormalities on other modalities (e.g. inter-ictal spikes). Some recent multimodal imaging analyses show that, for example, congruency/discrepancy between two modalities can in itself be a predictor of the epileptogenic zone, or good surgical outcome (e.g. Ridley *et al.* 2017, Coan *et al.* 2016).

Model inversion techniques also offer an alternative way to integrate multimodal data. Model inversion can be used to estimate model parameters given an empirical time series and a computational model capable of reproducing features of the signal. In the context of epilepsy, several studies have used model inversion to explore possible physiological bases of epileptic activity (Papadopoulou *et al.* 2015, Papadopoulou *et al.* 2017, Rosch *et al.* 2018, Karoly *et al.* 2018, Aarabi *et al.* 2014, Wendling *et al.* 2005, Freestone *et al.* 2014). However, these studies usually only model the activity of a localized area (or independently model the activity of each brain region), as model inversion of large, spatially coupled systems is typically computationally intractable. Although methods of model inversion for coupled systems are under development (Freestone *et al.* 2014), data from another modality (e.g. a dMRI derived network) may provide a way to constrain the coupling between brain regions (Stephan *et al.* 2009). Such models could potentially reveal the neural mechanisms underlying both local and global brain dynamics in epilepsy.

However, fully understanding how different modalities can inform personalised models will most likely require a more principled understanding of the relationship between modalities in the context of epilepsy. This will most likely entail mapping what the modalities really capture onto a biological basis, and relating them to each other, in the context of epilepsy. A related and similar argument also applies to different measures within the same modality. For example, structural connectivity strength can be defined in various ways (e.g. fractional anisotropy, streamline density, etc.), and a number of different measures have also been used to define functional connectivity. Different measures can lead to different predictions or validation outcomes. Again, a more principled understanding of different measures and their relationship to clinical variables is required.

4.2 Long-term longitudinal properties in epilepsy

Epilepsy symptoms change over timescales of years. For example, around 75% of patients are seizure-free after surgery for the first year. However, of those seizure-free patients, around one third will relapse to seizures again within five years (de Tisi et al, 2011). The long-term changes in seizure relapse/relief - and corresponding brain structure, dynamics and treatment impact - has barely been explored by the modelling community. The fact that patient retention is reasonably high means that longitudinal studies are possible. Models are ideal for studying dynamics regardless of the time scale; however, there are very few examples of disease progression modelling in epilepsy which are constrained by EEG or MRI (Abdelnour *et al.* 2015). Future studies should investigate long-timescale brain network changes and their relation to patient symptom changes, and could take inspiration from previous work in Alzheimer's research (Raj *et al.* 2012, Young *et al.* 2014).

Other than investigating long-term surgery outcome, the impact of medication can also be modelled by longitudinal personalised network models. To our knowledge, there are currently no such studies, which is surprising since anti-epileptic drugs (AEDs) are known to cause changes in the EEG (Wu & Ma 1993) and MRI properties (Pardoe *et al.* 2013). The lack of studies may be due to the lack of longitudinal data acquisition, and the fact that seizure frequency changes over time; therefore it is difficult to confirm if medication is responsible for any observed differences in dynamics. The ability to predict AED efficacy in new onset patients would be highly clinically valuable and network-based modelling offers opportunities to complement clinical decision making.

A related avenue of research that is currently not tackled in epilepsy is that of drug side-effect modelling, although outside of epilepsy, some general attempts in this direction are being made (Alberti *et al.* 2014). Drug side-effects have a huge impact on the quality of life in epilepsy patients, and are often linked to non-adherence and poor seizure management. It is conceivable to build personalised models to predict such side effects, although some systematic steps have to be made to quantify (ideally in an objective way) such side effects. Objective behavioural quantification methods (such as body sensor technologies) may open the door to this avenue in future.

4.3 Within-subject variability

While this chapter was concerned with building patient-specific models for prediction, one key aspect that has not been touched on is that of within-subject variability. For example, it is known that some patients may have different populations of seizures (Cook *et al.* 2016), some of which may be linked to specific brain states (e.g. awake vs. asleep - Bazil and Walczak 1997). Some populations may also be resistant to treatment. Even within a population, no two seizures ever look exactly alike. It has also been documented that seizures of shorter durations may preferentially appear together, as opposed to longer seizures in the same patient (Karoly *et al.* 2017). Similarly, short seizures appear to have a different underlying parameter evolution as compared to long seizures in the same patient (Karoly *et al.* 2018). This within-patient heterogeneity has serious consequences for treatment strategies. For example, surgical resection has to consider different epileptogenic zones for different seizure populations. Brain stimulation devices to prevent seizures, again, have to account for heterogeneous seizures within the same patient (Ewell *et al.* 2015).

Thus, it is also important to consider how personalised models can account for the within-subject variability across different brain states, times of day/week/months, and other modulatory processes. Essentially, we need to avoid overfitting our personalised models to particular (one-off) measurements, and be able to generalize them to "the patient", not just "the patient at one point in time". One way to begin to overcome this challenge is to understand the seizures in a patient as arising against a background of ongoing activity. The ongoing activity will modulate and influence how a seizure is going to occur and evolve (Khambhati *et al.* 2016). Hence understanding the interplay of the interictal state with the seizure mechanism through personalised models would be the next step forward.

In this chapter we have reviewed how dynamical systems models can be constrained by personalised patient data at the macroscopic scale with network interactions. We suggest that the development of models at multiple timescales (e.g. short-term seizure prediction and long-term disease progression) to predict and compare multiple treatment strategies is something the field of personalised modelling should strive toward.

5. Acknowledgements

We gratefully acknowledge funding from The Wellcome Trust (Seed award to YW 208940/Z/17/Z and Seed award to PNT 210109/Z/18/Z),

# 6. References

Aarabi, A. and He, B., 2014. Seizure prediction in hippocampal and neocortical epilepsy using a model-based approach. Clinical Neurophysiology, 125(5), pp.930-940.

Abdelnour, F., Mueller, S. and Raj, A., 2015. Relating cortical atrophy in temporal lobe epilepsy with graph diffusion-based network models. PLoS computational biology, 11(10), p.e1004564.

Ahmadi, M.E., Hagler, D.J., McDonald, C.R., Tecoma, E.S., Iragui, V.J., Dale, A.M. and Halgren, E., 2009. Side matters: diffusion tensor imaging tractography in left and right temporal lobe epilepsy. American journal of neuroradiology, 30(9), pp.1740-1747.

Alarcon, G., Binnie, C.D., Elwes, R.D.C. and Polkey, C.E., 1995. Power spectrum and intracranial EEG patterns at seizure onset in partial epilepsy. *Electroencephalography and clinical neurophysiology*, *94*(5), pp.326-337.

Alberti, P. and Cavaletti, G., 2014. Management of side effects in the personalised medicine era: chemotherapy-induced peripheral neuropathy. In Pharmacogenomics in Drug Discovery and Development (pp. 301-322). Humana Press, New York, NY.

Annegers, John F., et al. "A population-based study of seizures after traumatic brain injuries." *New England Journal of Medicine* 338.1 (1998): 20-24.

Baier, G., Goodfellow, M., Taylor, P.N., Wang, Y. and Garry, D.J., 2012. The importance of modeling epileptic seizure dynamics as spatio-temporal patterns. *Frontiers in physiology*, *3*, p.281.

Bazil, C.W. and Walczak, T.S., 1997. Effects of sleep and sleep stage on epileptic and nonepileptic seizures. Epilepsia, 38(1), pp.56-62.

Benjamin, O., Ashwin, P., Richardson, M. P., & Terry, J. R. (2012). The Journal of Mathematical Neuroscience A phenomenological model of seizure initiation suggests network structure may explain seizure frequency in idiopathic generalised epilepsy. The Journal of Mathematical Neuroscience, 2(1).

Besson, Pierre, Dinkelacker, Vera, Valabregue, Romain, Thivard, Lionel, Leclerc, Xavier, Baulac, Michel, Sammler, Daniela, et al., 2014. Structural connectivity differences in left and right temporal lobe epilepsy. NeuroImage 100, 135–144 (Elsevier).

Bonilha, Leonardo, Nesland, Travis, Martz, Gabriel U., Joseph, Jane E., Spampinato, Maria V., Edwards, Jonathan C., Tabesh, Ali, 2012. Medial temporal lobe epilepsy is associated with neuronal fibre loss and paradoxical increase in structural connectivity of limbic structures. J. Neurol. Neurosurg. Psychiatry 83 (9), 903–909

Burns, S.P., Santaniello, S., Yaffe, R.B., Jouny, C.C., Crone, N.E., Bergey, G.K., Anderson, W.S. and Sarma, S.V., 2014. Network dynamics of the brain and influence of the epileptic seizure onset zone. *Proceedings of the National Academy of Sciences*, *111*(49), pp.E5321-E5330.

Coan AC, Chaudhary UJ, Grouiller F, Campos BM, Perani S, De Ciantis A, Vulliemoz S, Diehl B, Beltramini GC, Carmichael DW, Thornton RC. EEG-fMRI in the presurgical evaluation of temporal lobe epilepsy. J Neurol Neurosurg Psychiatry. 2016 Jun 1;87(6):642-9.

Concha, L., Beaulieu, C. and Gross, D.W., 2005. Bilateral limbic diffusion abnormalities in unilateral temporal lobe epilepsy. Annals of neurology, 57(2), pp.188-196.

Cook, M. J., O'Brien, T. J., Berkovic, S. F., Murphy, M., Morokoff, A., Fabinyi, G., ... & Hosking, S. (2013). Prediction of seizure likelihood with a long-term, implanted seizure advisory system in patients with drug-resistant epilepsy: a first-in-man study. *The Lancet Neurology*, *12*(6), 563-571.

Cook, M.J., Karoly, P.J., Freestone, D.R., Himes, D., Leyde, K., Berkovic, S., O'brien, T., Grayden, D.B. and Boston, R., 2016. Human focal seizures are characterized by populations of fixed duration and interval. Epilepsia, 57(3), pp.359-368.
Da Silva, F.L., Blanes, W., Kalitzin, S.N., Parra, J., Suffczynski, P. and Velis, D.N., 2003. Epilepsies as dynamical diseases of brain systems: basic models of the transition between normal and epileptic activity. Epilepsia, 44, pp.72-83.

DeSalvo, Matthew N., Douw, Linda, Tanaka, Naoaki, Reinsberger, Claus, Stufflebeam, Steven M., 2014. Altered structural connectome in temporal lobe epilepsy. Radiology 270 (3), 842–848

De Tisi, Jane, Bell, Gail S., Peacock, Janet L., McEvoy, Andrew W., Harkness, William F.J., Sander, Josemir W., Duncan, John S., 2011. The long-term outcome of adult epilepsy surgery, patterns of seizure remission, and relapse: a cohort study. Lancet 378 (9800), 1388–1395

Duncan, John S., Gavin P. Winston, Matthias J. Koepp, and Sebastien Ourselin. "Brain imaging in the assessment for epilepsy surgery." *The Lancet Neurology* 15, no. 4 (2016): 420-433.

Ewell, L.A., Liang, L., Armstrong, C., Soltész, I., Leutgeb, S. and Leutgeb, J.K., 2015. Brain state is a major factor in preseizure hippocampal network activity and influences success of seizure intervention. Journal of Neuroscience, 35(47), pp.15635-15648.

Freestone, D.R., Karoly, P.J., Nešić, D., Aram, P., Cook, M.J. and Grayden, D.B., 2014. Estimation of effective connectivity via data-driven neural modeling. *Frontiers in neuroscience*, *8*, p.383.

Goodfellow, M., Rummel, C., Abela, E., Richardson, M.P., Schindler, K. and Terry, J.R., 2016. Estimation of brain network ictogenicity predicts outcome from epilepsy surgery. *Scientific reports*, *6*, p.29215.

Grinenko, O., Li, J., Mosher, J.C., Wang, I.Z., Bulacio, J.C., Gonzalez-Martinez, J., Nair, D., Najm, I., Leahy, R.M. and Chauvel, P., 2017. A fingerprint of the epileptogenic zone in human epilepsies. Brain, 141(1), pp.117-131.

Hagmann, P., Cammoun, L., Gigandet, X., Meuli, R., Honey, C.J., Wedeen, V.J. and Sporns, O., 2008. Mapping the structural core of human cerebral cortex. *PLoS biology*, *6*(7), p.e159.

Honey, C. J., & Sporns, O. (2008). Dynamical consequences of lesions in cortical networks. Human Brain Mapping, 29(7), 802–9.

Hutchings, F., Han, C. E., Keller, S. S., Weber, B., Taylor, P. N., & Kaiser, M. (2015). Predicting Surgery Targets in Temporal Lobe Epilepsy through Structural Connectome Based Simulations. PLoS Computational Biology, 11(12), 1–24.

Jansen, B. H., & Rit, V. G. (1995). Biological Cybernetics in a mathematical model of coupled cortical columns. *Biological Cybernetics*, *366*, 357–366.

Javidan, Manouchehr. "Electroencephalography in mesial temporal lobe epilepsy: a review." *Epilepsy research and treatment* 2012 (2012).

Jirsa, V.K., Stacey, W.C., Quilichini, P.P., Ivanov, A.I. and Bernard, C., 2014. On the nature of seizure dynamics. *Brain*, *137*(8), pp.2210-2230.

Jirsa, V.K., Proix, T., Perdikis, D., Woodman, M.M., Wang, H., Gonzalez-Martinez, J., Bernard, C., Bénar, C., Guye, M., Chauvel, P. and Bartolomei, F., 2017. The virtual epileptic patient: individualized whole-brain models of epilepsy spread. Neuroimage, 145, pp.377-388.

Karoly, Philippa J., Ewan S. Nurse, Dean R. Freestone, Hoameng Ung, Mark J. Cook, and Ray Boston. "Bursts of seizures in long-term recordings of human focal epilepsy." Epilepsia 58, no. 3 (2017): 363-372.

Karoly, P.J., Kuhlmann, L., Soudry, D., Grayden, D.B., Cook, M.J. and Freestone, D.R., 2018. Seizure pathways: A model-based investigation. *PLoS computational biology*, *14*(10), p.e1006403.

Khambhati, A.N., Davis, K.A., Lucas, T.H., Litt, B. and Bassett, D.S., 2016. Virtual cortical resection reveals push-pull network control preceding seizure evolution. Neuron, 91(5), pp.1170-1182.

Kramer, M.A., Eden, U.T., Kolaczyk, E.D., Zepeda, R., Eskandar, E.N. and Cash, S.S., 2010. Coalescence and fragmentation of cortical networks during focal seizures. Journal of Neuroscience, 30(30), pp.10076-10085.

Lopes, M.A., Richardson, M.P., Abela, E., Rummel, C., Schindler, K., Goodfellow, M. and Terry, J.R., 2017. An optimal strategy for epilepsy surgery: Disruption of the rich-club?. *PLoS computational biology*, *13*(8), p.e1005637.

Lossius, M.I., Hessen, E., Mowinckel, P., Stavem, K., Erikssen, J., Gulbrandsen, P. and Gjerstad, L., 2008. Consequences of antiepileptic drug withdrawal: a randomized, double-blind study (Akershus Study). Epilepsia, 49(3), pp.455-463.

Lu, J., Guo, S., Chen, M., Wang, W., Yang, H., Guo, D., & Yao, D. (2018). Generate the scale-free brain music from BOLD signals. *Medicine*, *97*(2).

Pacia, S.V. and Ebersole, J.S., 1997. Intracranial EEG substrates of scalp ictal patterns from temporal lobe foci. Epilepsia, 38(6), pp.642-654.

Papadopoulou, M., Leite, M., van Mierlo, P., Vonck, K., Lemieux, L., Friston, K. and Marinazzo, D., 2015. Tracking slow modulations in synaptic gain using dynamic causal modelling: validation in epilepsy. *Neuroimage*, *107*, pp.117-126.

Papadopoulou, M., Cooray, G., Rosch, R., Moran, R., Marinazzo, D. and Friston, K., 2017. Dynamic causal modelling of seizure activity in a rat model. *NeuroImage*, *146*, pp.518-532.

Proix, T., Bartolomei, F., Guye, M. and Jirsa, V.K., 2017. Individual brain structure and modelling predict seizure propagation. *Brain*, *140*(3), pp.641-654.

Merricks, E.M., Smith, E.H., McKhann, G.M., Goodman, R.R., Bateman, L.M., Emerson, R.G., Schevon, C.A. and Trevelyan, A.J., 2015. Single unit action potentials in humans and the effect of seizure activity. Brain, 138(10), pp.2891-2906.

Moeller, F., Siebner, H. R., Wolff, S., Muhle, H., Granert, O., Jansen, O., … Siniatchkin, M. (2008). Simultaneous EEG-fMRI in drug-naive children with newly diagnosed absence epilepsy. *Epilepsia*, *49*(9), 1510–9. http://doi.org/10.1111/j.1528-1167.2008.01626.x

Pardoe, Heath R., Anne T. Berg, and Graeme D. Jackson. "Sodium valproate use is associated with reduced parietal lobe thickness and brain volume." Neurology 80, no. 20 (2013): 1895-1900.

Proix, T., Jirsa, V. K., Bartolomei, F., Guye, M., & Truccolo, W. (2018). Predicting the spatiotemporal diversity of seizure propagation and termination in human focal epilepsy. *Nature communications*, *9*(1), 1088.

Raj, A., Kuceyeski, A. and Weiner, M., 2012. A network diffusion model of disease progression in dementia. Neuron, 73(6), pp.1204-1215.

Richardson, M.P., 2012. Large scale brain models of epilepsy: dynamics meets connectomics. *J Neurol Neurosurg Psychiatry*, pp.jnnp-2011.

Ridley, B., Wirsich, J., Bettus, G., Rodionov, R., Murta, T., Chaudhary, U., Carmichael, D., Thornton, R., Vulliemoz, S., McEvoy, A. and Wendling, F., 2017. Simultaneous intracranial EEG-fMRI shows inter-modality correlation in time-resolved connectivity within normal areas but not within epileptic regions. Brain topography, 30(5), pp.639-655.

Rosch, R.E., Wright, S., Cooray, G., Papadopoulou, M., Goyal, S., Lim, M., Vincent, A., Upton, A.L., Baldeweg, T. and Friston, K.J., 2018. NMDA-receptor antibodies alter cortical microcircuit dynamics. *Proceedings of the National Academy of Sciences*, *115*(42), pp.E9916-E9925.

Schmidt, H., Petkov, G., Richardson, M.P. and Terry, J.R., 2014. Dynamics on networks: the role of local dynamics and global networks on the emergence of hypersynchronous neural activity. PLoS computational biology, 10(11), p.e1003947.

Sinha N, Dauwels J, Wang Y, Taylor P. An in silico approach for pre-surgical evaluation of an epileptic cortex. IEEE Proc EMBS 2014

Sinha, N., Dauwels, J., Kaiser, M., Cash, S. S., Brandon Westover, M., Wang, Y., & Taylor, P. N. (2017). Predicting neurosurgical outcomes in focal epilepsy patients using computational modelling. Brain : A Journal of Neurology, 140(Pt 2), 319–332.

Smith, E.H. and Schevon, C.A., 2016. Toward a mechanistic understanding of epileptic networks. Current neurology and neuroscience reports, 16(11), p.97.

Stephan, K.E., Tittgemeyer, M., Knösche, T.R., Moran, R.J. and Friston, K.J., 2009. Tractography-based priors for dynamic causal models. *Neuroimage*, *47*(4), pp.1628-1638.

Taussig, D., A. Montavont, and J. Isnard. "Invasive EEG explorations." *Neurophysiologie Clinique/Clinical Neurophysiology* 45.1 (2015): 113-119.

Taylor PN, Goodfellow M, Wang Y, Baier G. Towards a large-scale model of patient-specific epileptic spike-wave discharges. Biol Cybern (2013) 107:83–94.

Taylor, P., Baier, G., Cash, S., & Dauwels, J. (2013b). A model of stimulus induced epileptic spike-wave discharges. In IEEE CCMB (pp. 1–7).

Taylor, P. N., Kaiser, M., & Dauwels, J. (2014). Structural connectivity based whole brain modelling in epilepsy. *Journal of Neuroscience Methods*, *236*, 51–57.

Taylor, P. N., Thomas, J., Sinha, N., Dauwels, J., Kaiser, M., Thesen, T., & Ruths, J. (2015). Optimal control based seizure abatement using patient derived connectivity. Frontiers in Neuroscience, 9(June), 1–10.

Taylor, P.N., Sinha, N., Wang, Y., Vos, S.B., de Tisi, J., Miserocchi, A., McEvoy, A.W., Winston, G.P. and Duncan, J.S., 2018. The impact of epilepsy surgery on the structural connectome and its relation to outcome. *NeuroImage: Clinical*, *18*, pp.202-214.

Truccolo, W., Donoghue, J.A., Hochberg, L.R., Eskandar, E.N., Madsen, J.R., Anderson, W.S., Brown, E.N., Halgren, E. and Cash, S.S., 2011. Single-neuron dynamics in human focal epilepsy. Nature neuroscience, 14(5), p.635.

Weiss, S.A., Lemesiou, A., Connors, R., Banks, G.P., McKhann, G.M., Goodman, R.R., Zhao, B., Filippi, C.G., Nowell, M., Rodionov, R. and Diehl, B., 2015. Seizure localization using ictal phase-locked high gamma A retrospective surgical outcome study. Neurology, 84(23), pp.2320-2328.

Wendling, F., Bartolomei, F., Bellanger, J. J., & Chauvel, P. (2002). Epileptic fast activity can be explained by a model of impaired GABAergic dendritic inhibition. Eur. J of Neuroscience, 15, 1499–1508.

Wendling, F., Hernandez, A., Bellanger, J. J., Chauvel, P., & Bartolomei, F. (2005). Interictal to ictal transition in human temporal lobe epilepsy: insights from a computational model of intracerebral EEG. Journal of Clinical Neurophysiology, 22(5), 343.

Wilson, H.R. and Cowan, J.D., 1972. Excitatory and inhibitory interactions in localized populations of model neurons. Biophysical journal, 12(1), pp.1-24.

Winston, G.P., Yogarajah, M., Symms, M.R., McEvoy, A.W., Micallef, C. and Duncan, J.S., 2011. Diffusion tensor imaging tractography to visualize the relationship of the optic radiation to epileptogenic lesions prior to neurosurgery. *Epilepsia*, *52*(8), pp.1430-1438.

Winston, G.P., Daga, P., Stretton, J., Modat, M., Symms, M.R., McEvoy, A.W., Ourselin, S. and Duncan, J.S., 2012. Optic radiation tractography and vision in anterior temporal lobe resection. *Annals of neurology*, *71*(3), pp.334-341.

Wu, X., and J. J. Ma. "Sodium valproate: Quantitative EEG and serum levels in volunteers and epileptics." *Clinical Electroencephalography* 24.2 (1993): 93-99.

Yan B, Li P. The emergence of abnormal hypersynchronization in the anatomical structural network of human brain. NeuroImage 2013;65:34–51

Yang, C., Luan, G., Wang, Q., Liu, Z., Zhai, F. and Wang, Q., 2018. localization of epileptogenic Zone With the correction of Pathological networks. Frontiers in neurology, 9, p.143.

Young, A.L., Oxtoby, N.P., Daga, P., Cash, D.M., Fox, N.C., Ourselin, S., Schott, J.M. and Alexander, D.C., 2014. A data-driven model of biomarker changes in sporadic Alzheimer's disease. Brain, 137(9), pp.2564-2577.

Zheng, T. W., O'Brien, T. J., Morris, M. J., Reid, C. a, Jovanovska, V., O'Brien, P., … Pinault, D. (2012). Rhythmic neuronal activity in S2 somatosensory and insular cortices contribute to the initiation of absence-related spike-and-wave discharges. Epilepsia, 1–11